\RequirePackage{fix-cm}
\documentclass[smallextended]{svjour3} 
\smartqed 

\usepackage{eso-pic}
\usepackage{graphicx}
\usepackage{booktabs}
\usepackage{marginnote}

\newcommand{\tmu}{\textmu}
\newcommand{\sss}{\scriptscriptstyle}
\newcommand{\sst}{\scriptstyle}

\newcommand{\stext}[1]{\sss \rm{#1} \sst}

\usepackage{marginnote}

\usepackage{eso-pic}

\begin{document}
\emergencystretch 3em
\title{Mechanical Tuning of the Terahertz Photonic Bandgap of 3D-Printed One-Dimensional Photonic Crystals}
\author{Serang Park$^{1}$\and
        Brandon Norton$^{1}$\and
        Glenn D.~Boreman$^{1}$\and        
        Tino Hofmann$^{1}$ }
\institute{Serang Park \at
\email{spark71@uncc.edu}\\
Phone: +1 704 451 6937 \\
              1) Department of Physics and Optical Science, University of North Carolina at Charlotte, 9201 University City Blvd, Charlotte, NC, 28223\at
}
\date{Received: date / Accepted: date}
\maketitle

\begin{abstract}

Mechanical tuning of a 3D-printed, polymer-based one-dimensional photonic crystal was demonstrated in the terahertz spectral range. The investigated photonic crystal consists of 13 alternating compact and low-density layers and was fabricated through single-step stereolithography. While the compact layers are entirely polymethacrylate without any intentional internal structures, the low-density layers contain sub-wavelength sized slanted columnar inclusions to allow the mechanical compression in a direction normal to the layer interfaces of the photonic crystal. Terahertz transmission spectroscopy of the photonic crystal was performed in a spectral range from 83 to 124~GHz as a function of the compressive strain. The as-fabricated photonic crystal showed a distinct photonic bandgap centered at 109~GHz, which blue shifted under compressive stress. A maximum shift of 12~GHz in the bandgap center frequency was experimentally demonstrated. Stratified optical models incorporating simple homogeneous and inhomogeneous compression approximations were used to analyze the transmission data. A good agreement between the experimental and model-calculated transmission spectra was found.

\end{abstract}

\section{Introduction}
\label{intro}

Additive manufacturing techniques have recently gained attention as tools for direct fabrication of classical optical components including lenses~\cite{Squires20153D,Busch2016THz,Hernandez2016Fabrication}, filters~\cite{Kaur2015Affordable,ortiz2019guided}, waveguides and diffractive optics~\cite{Yang20163D,Furlan20163D} for the terahertz (THz) spectral range. While the majority of the research is focused on fused-filament deposition approaches, stereolithography has been demonstrated as a viable alternative in particular for applications where low surface roughness and high spatial resolution are required~\cite{Fullager2019Metalized,Park2019Terahertz}. State-of-the-art commercial stereolithography systems can achieve spatial resolutions on the order of 10~\tmu m and therefore allow the synthesis of functional THz optical components with significantly improved surface roughness compared to other additive manufacturing techniques~\cite{Ngo2018Additive}. The spatial resolution achieved with stereolithography systems further enables the fabrication of  polymethacrylate-based, one-dimensional photonic crystals for the THz spectral range~\cite{Park2020one}.

Photonic crystals are composed of periodically arranged sub-wavelength-sized constituents, which influence propagation of electromagnetic radiation by forming photonic bandgaps~\cite{Yablonovitch1987Inhibited}. Propagation of electromagnetic radiation with energies within the photonic bandgap is prohibited, resulting in spectral regions with high reflectivity~\cite{Joannopoulos1997Photonic}. The characteristics of the photonic bandgap, such as spectral position and width, depend on the geometry of the unit cell and the dielectric properties of the photonic crystal~\cite{Yablonovitch1987Inhibited}. This dependency can be utilized to achieve dynamically tunable photonic crystals through thermal~\cite{Honda2009dual,Nemec2004Thermally} and electrical stimuli~\cite{Li2003Ferroelectric,Xia2005electric}. 

Common to these tuning approaches is to change the dielectric function of the photonic crystal constituents, which results in a change of the energy of the photonic bandgap~\cite{Li2011tunable}. Mechanical tuning of photonic crystals has also been studied, in which geometrical changes of the photonic crystal structure are primarily exploited~\cite{Park2004mechanically}. While a large body of literature exists on mechanically tunable photonic crystals designed for the infrared and visible spectral range, e.g.~\cite{Sumioka2002tuning,Kimura1979tunable,Yoshino1999mechanical}, mechanical tuning of photonic crystals designed for the THz spectral range has not yet been widely explored~\cite{Zhu2017direct}. 

In this paper, we demonstrate the mechanical tuning of stereolithographically fabricated one-dimensional photonic crystals in the THz frequency range. We achieve the maximum shift of 12~GHz in the center frequency of the photonic bandgap by applying compressive stress normal to the layer interfaces of the photonic crystal. Stratified optical layer calculations are used to analyze THz transmission spectra obtained at normal incidence in the range from 83 to 124~GHz.

\section{Experiment}
\label{sec:experiment}

\subsection{Design and Fabrication}
\label{sec:design and fabrication}

\begin{figure}[htb]
	\centering
	\includegraphics[width=1\linewidth, keepaspectratio=true, trim=30 30 35 35, clip]{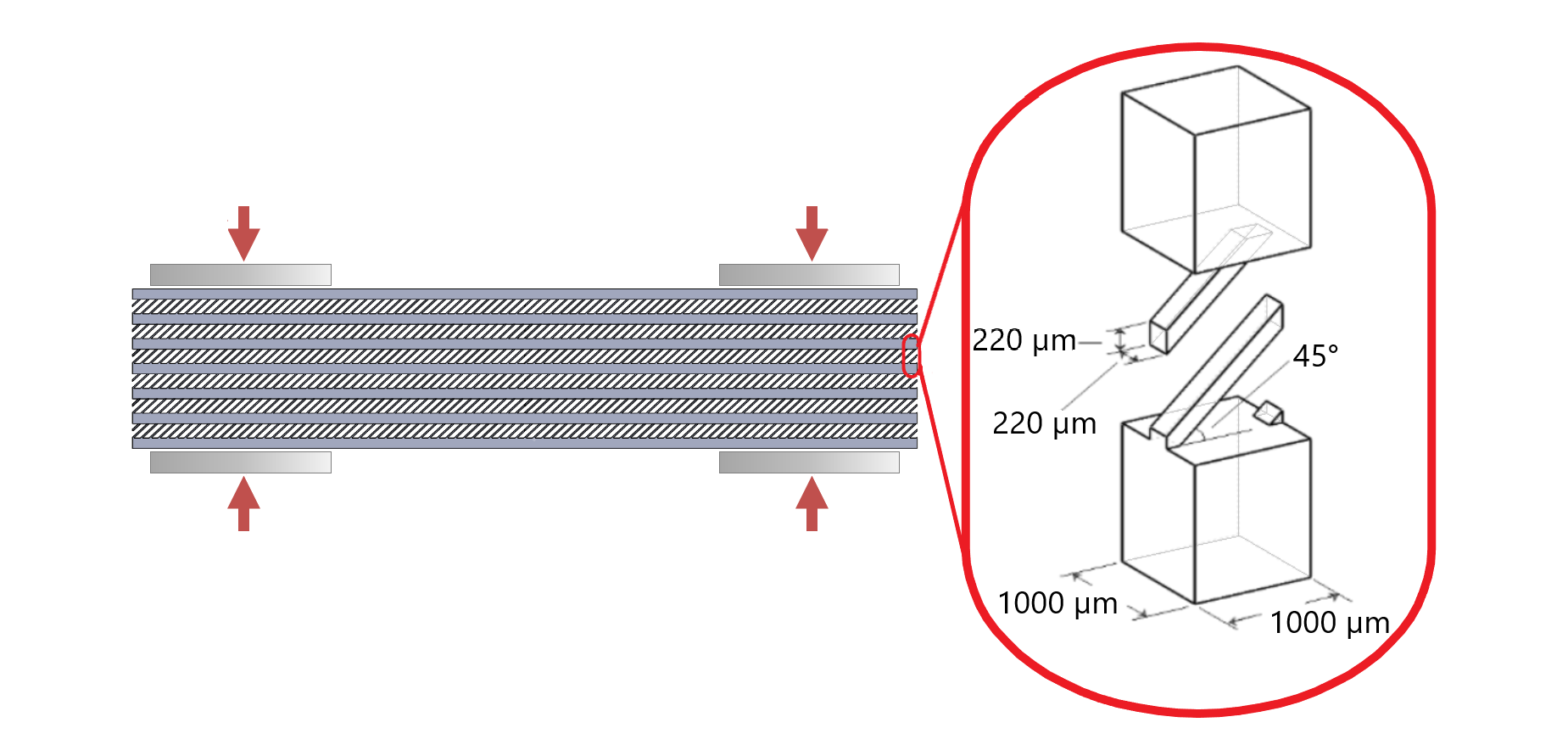}
    \caption{A side view schematic of the photonic crystal composed of 13 alternating compact and low-density layers. The direction of the external force is normal to the layer interfaces as indicated by the red arrows. The low-density layers are composed of columnar structures oriented at 45$^{\circ}$ with respect to the layer interfaces and arranged in square lattice pattern as shown in the inset. The slanting plane is perpendicular to the interface of the layers.}
    \label{designfig}
\end{figure}

The photonic crystals studied here are composed of six low-density and seven compact layers alternating throughout the structure. The alternating layers have plane parallel interfaces and were fabricated through single step stereolithography, using a single polymethacrylate which is semitransparent in the THz spectral range~\cite{Park2019Terahertz}. The compact layers consist of polymethacrylate without any intentional internal structure. The low-density layers are composed of sub-wavelength sized columnar structures with a square base oriented at 45$^{\circ}$ with respect to the layer interfaces. A schematic of the photonic crystal including the unit cell is depicted in Fig.~\ref{designfig}. The orientation of the columnar structures was chosen so that the low-density layers are compressible when subjected to an external force normal to the layer interfaces as indicated by the arrows in Fig.~\ref{designfig}.

Stratified optical layer model calculations using the commercial software package (WVASE32\textsuperscript{TM}, J.A.~Woollam Co., Inc.) were employed to determine an optimal design of a photonic crystal with a bandgap located in the F-band with a center frequency of approximately 112~GHz. Optical model calculations for the design shown in Fig.~\ref{designfig} only require the knowledge of the dielectric function of the compact layers as the dielectric function of the low-density layers can be obtained using the Bruggeman effective medium approximation \cite{Park2020one,Li2018High}. The dielectric function of the polymethacrylate (black v4, Formlabs Inc.), which was used for the fabrication of the photonic crystals investigated here, is well known and was determined using spectroscopic ellipsometry in the infrared and THz spectral range \cite{Park2019Terahertz}. The dielectric function for the low-density layers $\varepsilon_{\stext{l}}^{\stext{eff}}$ is described using the Bruggeman effective medium approximation as the columnar structures shown in Fig.~\ref{eqn:Bruggeman} are sufficiently small compared to the wavelength of the incident THz radiation. 

$\varepsilon_{\stext{l}}^{\stext{eff}}$ can be expressed as a function of the volumetric fraction of the columnar inclusions $f_{\stext{i}}$, the dielectric function of the columnar inclusions $\varepsilon_{\stext{i}}$, which in our case is the dielectric function of the polymethacrylate, and the dielectric function of the host $\varepsilon_{\stext{h}}$, air~\cite{Cai2010optical}:
\begin{eqnarray}
\label{eqn:Bruggeman}
\varepsilon_{\stext{l}}^{\stext{eff}}(f_{\stext{i}})&=&\frac{1}{4}\Big\{(3f_{\stext{i}}-1)\varepsilon_{\stext{i}}+(2-3f_{\stext{i}})\varepsilon_{\stext{h}}\pm \\\nonumber	&&\sqrt{[(3f_{\stext{i}}-1)\varepsilon_{\stext{i}}+(2-3f_{\stext{i}})\varepsilon_{\stext{h}}]^2+8\varepsilon_{\stext{i}}\varepsilon_{\stext{h}}}\Big\}.
\end{eqnarray}

For slanted columnar inclusions with a square cross section arranged in a square lattice pattern as shown in Fig.~\ref{designfig}, the volumetric fraction of the inclusions $f_{\stext{i}}$ can be expressed as a function of the thickness of the low-density layer $d_{\stext{l}}$ and a coefficient $f_0$, derived from the width of the columnar structures $w_{\stext{c}}$, the width of the unit cell $w_{\stext{u}}$, and the length of the columnar structures $l_{\stext{c}}$: 
\begin{equation}
\label{eqn:fraction}
f_{\stext{i}}(d_{\stext{l}})=\frac{w_{\stext{c}}^2l_{\stext{c}}}{w_{\stext{u}}^2}\frac{1}{d_{\stext{l}}}=f_0\frac{1}{d_{\stext{l}}}.
\end{equation}

\noindent Thus, $\varepsilon_{\stext{l}}^{\stext{eff}}$ presented in Eqn.~(\ref{eqn:Bruggeman}) is a function of the low-density layer thickness $d_{\stext{l}}$. The optimized nominal values to achieve a photonic bandgap centered at 112~GHz are $f_{\stext{i}}=0.05$, $d_{\stext{l}}=1800$~\tmu m, and $d_{\stext{c}}=1300$~\tmu m.

\begin{figure}[htb]
	\centering
	\includegraphics[width=1\linewidth, keepaspectratio=true, trim=0 0 0 25, clip]{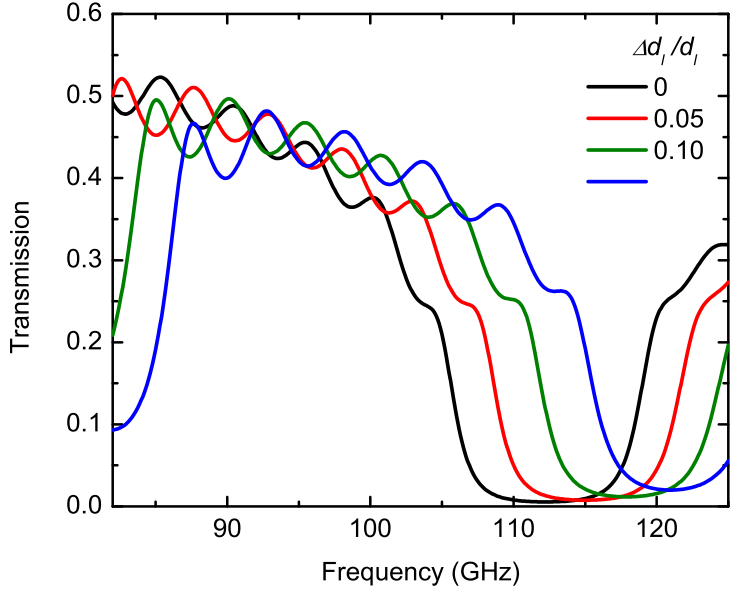}
    \caption{Model calculated spectra of the photonic crystal for compressive strain values in the low-density layers, $\Delta d_{\stext{l}} / d_{\stext{l}}=0$, 0.05, 0.10, and 0.15. The bandgap center frequency is blue shifted and the minimum transmission slightly increases with increasing compressive strain.}
    \label{nominalfig}
\end{figure}

Figure \ref{nominalfig} shows the calculated transmission spectra for the nominal photonic crystal design in the range from 82 to 125~GHz for several different compressive strain values $\Delta d_{\stext{l}} / d_{\stext{l}}=0$, 0.05, 0.10, and 0.15. It can be clearly seen that the  photonic bandgap, which is centered around 112~GHz for $\Delta d_{\stext{l}} / d_{\stext{l}}=0$, shifts to higher frequencies as the compressive strain is increased. For $\Delta d_{\stext{l}} / d_{\stext{l}}=0.15$, a center frequency of 120~GHz can be observed. The minimum transmission of the photonic bandgap also slightly increases with increasing compressive strain. For the calculations, a homogeneous compression model was implemented. In this model, the low-density layers are assumed to be homogeneously compressed and thus their dielectric function can be expressed as a function of the thickness following Eqns.~(\ref{eqn:Bruggeman}) and (\ref{eqn:fraction}). 

The photonic crystal structure was designed using a 3D CAD software (SolidWorks, Dassault Systèmes) and fabricated using a commercial 3D printer (Form 2, Formlabs Inc.). The printer uses an inverted top-down approach, employing a 405~nm UV laser to polymerize a photosensitive resin (black v4, Formlabs Inc.) onto a printing platform layer by layer. This requires support structures to keep the printed object fixed to the printing platform, which allows the object to be printed in an arbitrary position. The photonic crystals investigated here were oriented at a 45$^\circ$ angle relative to the normal of the printing platform to prevent any unpolymerized resin from accumulating within the structure throughout the printing process. After the photonic crystal was printed, all support structures were removed and the sample was placed in an isopropanol bath for 20 minutes. It was then thoroughly rinsed with isopropanol for additional 10 minutes before it was dried using pressurized air to remove any excess resin or isopropanol from the low-density layers.

\subsection{Data Acquisition and Analysis}

An electronic synthesizer (Virginia Diodes Inc.) connected to an extension module (Virginia Diodes Inc.) was used as a source. The emitted radiation was collimated using a 60~mm focal length lens before being transmitted through the photonic crystal sample. The beam transmitted through the photonic crystal was then focused into a broadband power meter (PM3, Erickson Instruments) using a second 60~mm focal length lens and the power was recorded as a function of frequency and compressive strain, which was measured using a micrometer caliper. Figure~\ref{designfig} illustrates how the photonic crystal was compressed to mechanically tune its optical response. The direction of the compressive stress is indicated by red arrows. The photonic crystal was orientated such that the polarization direction of the source radiation is parallel to the slanting plane of the columnar structures. This high-symmetry orientation avoids polarization mode conversion due to form-birefringence caused by the slanted columnar structures, which were in Ref.~\cite{park2020terahertz}.
The spectral response of the photonic crystal was measured for two compressive strain values $\Delta d_{\stext{l}}/d_{\stext{l}} =$ 0.18$\pm$0.01, and 0.21$\pm$0.01 in the range from 83 to 124~GHz with a resolution of 0.1~GHz. 

\begin{figure}[htb]
	\centering
	\includegraphics[width=1\linewidth, keepaspectratio=true, trim=20 130 160 0, clip]{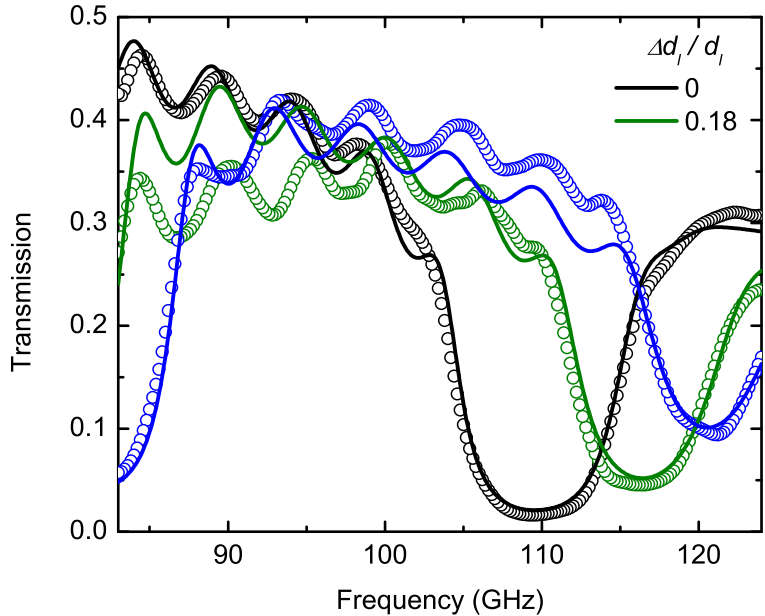}
    \caption{Experimental (circles) and best-model calculated (solid red lines) transmission spectra of the photonic crystal for different compressive strain values, $\Delta d_{\stext{l}}/d_{\stext{l}}$, in the spectral range from 83 to 124~GHz. Compressing the crystal results in a blue shift of the bandgap's center frequency. The bandgap center frequencies for $\Delta d_{\stext{l}}/d_{\stext{l}} = 0,$ 0.18, and 0.21 are found to be 109~GHz, 116~GHz, and 121~GHz with the minimum transmission of 0.02, 0.05, and 0.1, respectively.}
    \label{compressionfig}
\end{figure}

\begin{figure}[htb]
	\centering
	\includegraphics[width=1\linewidth, keepaspectratio=true, trim=0 0 0 0, clip]{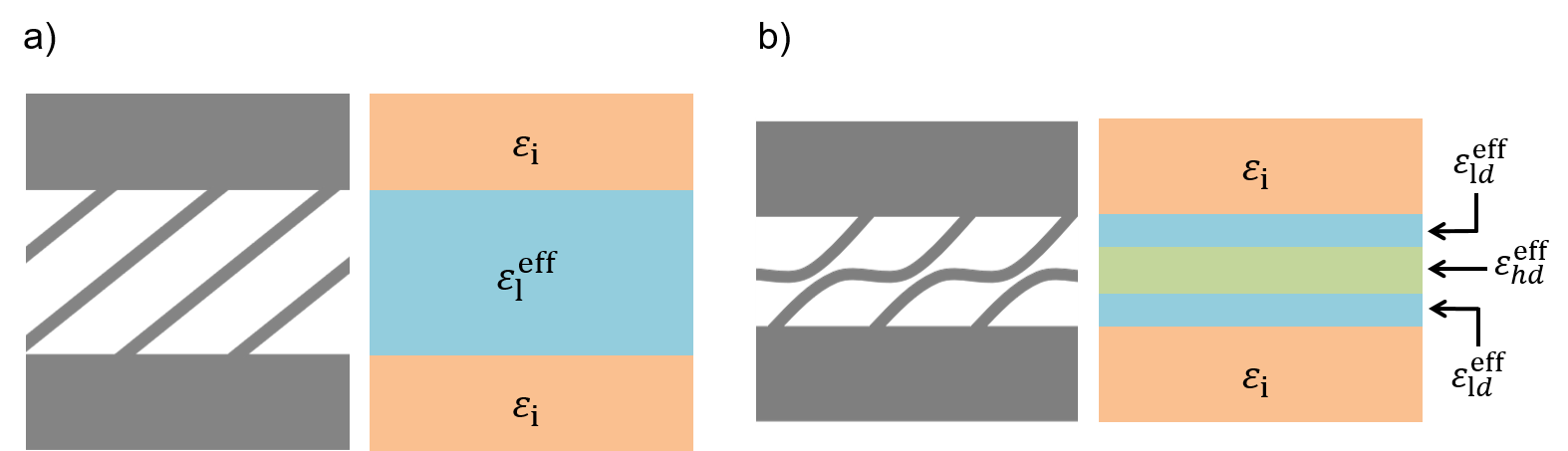}
    \caption{Schematic of a set of low-density and compact layers (side view) before a) and after compression b). The layer optical model is shown in comparison. While the low-density and compact layers of the un-strained photonic crystal can be described in the optical model as homogeneous thin films with a thickness $d_{\stext{l}}$ and $d_{\stext{c}}$, respectively, this model breaks down for the strained samples. An accurate description of the experimental results needs to account for the experimental inhomogeneous compression. The low density layers are therefore approximated by three layers ($d_{\stext{ld}}$, $d_{\stext{hd}}$) in the optical model. The dielectric functions of these layers are denoted by $\varepsilon^{\stext{eff}}_{\stext{ld}}$ and $\varepsilon^{\stext{eff}}_{\stext{hd}}$.}
    \label{analysisfig}
\end{figure}

The experimental THz transmission data shown in Fig.~\ref{compressionfig} were analyzed using stratified layer model calculations employing a commercial software package (WVASE32\textsuperscript{TM}, J.A.~Woollam Company). The optical model used to analyze the un-strained, as fabricated photonic crystal is composed of 13 layers alternating between compact and low-density material as shown in comparison with a schematic of a section of the photonic crystal in Fig.~\ref{analysisfig}~a). A Levenberg–Marquardt algorithm is used during the analysis to vary the optical model parameters until the best-match between experimental and model calculated data is achieved. For the as-fabricated photonic crystal the thickness of the constituent layers were varied during the analysis. It was assumed that the dielectric function and layer thickness of the compact and low-density layers is constant throughout the photonic crystal. Therefore, the corresponding parameters of the optical model $d_{\stext{c}}$, $d_{\stext{l}}$, and $\varepsilon^{\stext{eff}}_{\stext{l}}$ were varied for all layers simultaneously.       

The simple homogeneous compression model introduced in Sect.~\ref{sec:design and fabrication} is insufficient to describe the experimental transmission spectra obtained for the photonic crystal under compressive stress (see Fig.~\ref{compressionfig}). Therefore a simple form of an inhomogeneous compression is implemented in the optical model used to analyze the experimental transmission data obtained for the photonic crystal under compressive stress. For this model, which is shown schematically in Fig.~\ref{analysisfig}~b), each low-density layer consists of three layers where the low-density layers adjacent to compact layers are treated as incompressible. Thus the resulting three low-density layers resemble a very simple approximation of the experimentally observed compression gradient. The low-density layer thickness, the compact layer thickness, and the volume fraction of the columnar inclusions were varied during this model analysis.

\section{Results and Discussion}

Figure~\ref{compressionfig} shows the experimental (circles) and best-model (red solid lines) transmission spectra in the spectral range from 83 to 124~GHz and compressive strain $\Delta d_{\stext{l}} / d_{\stext{l}}$ values of 0, 0.18, and 0.21 at normal incidence. All spectra show a distinct photonic bandgap and typical sidelobes. It can be clearly seen that the photonic bandgap shifts to higher frequencies as a function of the compressive strain $\Delta d_{\stext{l}} / d_{\stext{l}}$. In addition to the spectral shift of the main photonic band gap, a secondary photonic bandgap is shifting into experimentally accessible spectral window for the largest strain $\Delta d_{\stext{l}} / d_{\stext{l}}=0.21$.

The best-model calculated transmission spectra render the experimentally observed response well. In comparison to the calculated transmission spectra obtained with a homogeneous compression model as shown in Fig.~\ref{nominalfig}, it can be noted that the increase in the transmission minimum and the change in the shape of the photonic bandgap is well reproduced in the inhomogeneous compression model-based calculated spectra. All  experimental transmission spectra were analyzed simultaneously using optical models for which common parameters ($\varepsilon_{\stext{i}}$, $\varepsilon^{\stext{eff}}_{\stext{ld}}$, $d_{\stext{c}}$, and $d_{\stext{ld}}$) were not varied independently.

For the un-strained, as-fabricated photonic crystal the model analysis yields $d_{\stext{c}}$=1426$\pm$3~\tmu m for the thickness of the compact layers and $d_{\stext{l}}$=1659$\pm$5~\tmu m for the thickness of the low-density layers. The best-fit parameters found for the fabricated crystal deviate slightly from the nominal values. We attribute this small deviation to the resolution of the stereolithography system employed here, which has a nominal layer resolution of 25~\tmu m in the direction orthogonal to the polymerization plane and a nominal laser spot size of 140~\tmu m \cite{park2020terahertz}. 

For a compressive strain of $\Delta d_{\stext{l}}/d_{\stext{l}}=0.18\pm0.01$, the center frequency of the photonic bandgap is shifted by 7~GHz to 116~GHz, while the transmission minimum increases to 0.05. The best-model parameters for the thicknesses of the low density layers is $d_{\stext{ld}}=$ 332$\pm$10~\tmu m and $d_{\stext{hd}}= $714$\pm$22~\tmu m. The best-model value for the compressive strain $\Delta d_{\stext{l}}/d_{\stext{l}}=0.17\pm 0.03$, is found to be in good agreement with the value determined mechanically using a micrometer caliper. 

For a compressive strain of $\Delta d_{\stext{l}}/d_{\stext{l}}=0.21\pm0.02$ the center frequency of the photonic bandgap is shifted by 12~GHz to 121~GHz, while the transmission minimum increases to 0.05. In addition, the high-frequency flank of a second photonic bandgap is emerging at the lower end of the experimentally accessible spectral window. The best-model parameters for the thicknesses of the low density layers is $d_{\stext{ld}}=$ 332$\pm$10~\tmu m and $d_{\stext{hd}}= $574$\pm$22~\tmu m. The best-model value for the compressive strain $\Delta d_{\stext{l}}/d_{\stext{l}}=0.25\pm 0.03$. Similar to the lower compressive strain this values is also found in good agreement with the mechanically determined value.

\section{Conclusion}
\label{conclusion}

A mechanically tunable, one-dimensional THz photonic crystal fabricated using stereolithography has been demonstrated for the F-band. The photonic crystal is composed of 13 alternating compact and low-density layers. The compact layers consist of polymethacrylate without any intentional internal structure. The low-density layers are composed of sub-wavelength sized columnar structures that are oriented at 45$^\circ$ relative to the layer interfaces. This slanted columnar design allows the mechanical compression in a direction normal to the layer interfaces. 

THz transmission spectroscopy of the photonic crystal was performed in a spectral range from 83 to 124~GHz at normal incidence as a function of the applied compressive stress. The as-fabricated photonic crystal shows a distinctive photonic bandgap centered at 109~GHz. Under compressive stress, applied perpendicular to the layer interfaces, the photonic bandgap shifts to higher frequencies. The largest observed shift is 12~GHz for a compressive strain of $\Delta d_{\stext{l}}/d_{\stext{l}}=0.21\pm0.02$. 

Stratified layer model calculations were used to analyze the THz transmission spectra. An optical model composed of 13 alternating compact and low-density layers. The effective dielectric function of the low-density layers was described through Bruggeman effective medium approximation. This model was found to render the experimental spectrum obtained for as-fabricated photonic crystal accurately. However, this model was found to be insufficient to describe the THz transmission data obtained for photonic crystals subjected to different compressive stresses. Spectral signatures indicative of inhomogeneous compression were observed and led to the development of a simple inhomogeneous compression model, which describes the experimentally observed transmission spectra well. 

In conclusion, a mechanically tunable photonic crystal for the THz spectral range is demonstrated using a simple stereolithographic fabrication. We envision a wide range of THz applications including surveillance and sensing for such a structure. Our design allows the mechanical tuning of the photonic bandgap within the F-band.

\begin{acknowledgements}
	SP, BN, and TH would like to acknowledge the valuable discussions with Susanne Lee (L3Harris Technologies, Inc.) and Stefan Sch\"oche (J.A.~Woollam Co., Inc.) within the NSF IUCRC for Metamaterials. The authors are grateful for support from the National Science Foundation (1624572) within the IUCRC Center for Metamaterials, the Army Research Office
(W911NF-14-1-0299), and the Department of Physics and Optical Science of the University of North Carolina at Charlotte.
\end{acknowledgements}

\end{document}